\documentclass[a4paper,11pt]{article}
\usepackage{pos}
\renewcommand{\logo}{\relax}
\renewcommand{\hookBeforeTitle}{\hfill\textsc{BU-HEPP-22-04}
	\vskip2em   plus .4fill  minus 1em}
\def\rQCED{{\rm QCED}}
\newcommand{\KK}{{KK}}
\title{Overview of IR-Improvement in Precision LHC$/$FCC Physics}

\author*[a]{S. A. Yost}
\author[b]{Y. Liu}
\author[b]{B. Shakerin}
\author[b]{B. F. L. Ward}

\affiliation[a]{Department of Physics, The Citadel,\\
  171 Moultrie St., Charleston, South Carolina 29409, USA}

\affiliation[b]{Department of Physics, Baylor University,\\
One Bear Place 97316, Waco, Texas 76798, USA}

\emailAdd{Scott.Yost@citadel.edu}
\emailAdd{Yang\_Liu1@baylor.edu}
\emailAdd{Bahram.Shakerin@gmail.com}
\emailAdd{BFL\_Ward@baylor.edu}

\abstract{We present an overview of the use of IR-improvement of unintegrable singularities in the infrared regime via amplitude-based resummation in $\mathrm{QED} \times \mathrm{QCD}$ $\subset$ $\mathrm{SU}(2)_L \times \mathrm{U}_1 \times \mathrm{SU}(3)^c$. We work in the context of precision LHC/FCC physics. While illustrating such IR-improvement in specific examples, we discuss new results and new issues.}

\FullConference{%
  41st International Conference on High Energy physics - ICHEP2022\\
  6-13 July, 2022\\
  Bologna, Italy
}

\begin{document}
\maketitle

\section{Introduction}
The LHC and the planned FCC~\cite{fcc-study} both require precision theory 
predictions to exploit fully their physics potentials. One of the theoretical 
tools that is key to precision theory is resummation in the context of quantum 
field theory.\footnote{See Ref.~\cite{scipost-ward-2021} for more discussion 
on the various types of resummation in the latter context.} 
Refs.~\cite{hwri,mcnlo-hwri,kkmchh23,kkmchh4} extend the original exact 
resummation theory of Yennie, Frautschi, and Suura~\cite{yfs61}
to an amplitude-level exact resummation for the entire 
QCD$\otimes$EW theory, resulting in IR-improvement of the 
comparison between data and the respective Standard Model predictions. 
Here, we give a brief 
report on the status of this IR-improvement program, beginning with a concise
review of exact amplitude-based resummation methods and following with some
new results and issues in precision LHC and FCC physics.\par
\section{Concise Review of Exact Amplitude-Based Resummation Theory}
 The application of exact amplitude-based resummation theory to precision LHC/FCC physics begins with the master formula for the process $q_1+q_2 \rightarrow q'_1+q'_2$, in an obvious notation for the kinematics,
\begin{eqnarray}
&d\bar\sigma_{\rm res} = e^{\rm SUM_{IR}(QCED)}
   {\sum}_{{n,m}=0}^\infty\frac{1}{n!m!}\int\prod_{j_1=1}^n\frac{d^3k_{j_1}}{k_{j_1}} 
\frac{d^3p_2}{p_2^{\,0}}\frac{d^3q_2}{q_2^{\,0}}
\prod_{j_2=1}^m\frac{d^3k'_{j_2}}{k'_{j_2}}
\cr &\qquad
\int\frac{d^4y}{(2\pi)^4}e^{iy\cdot(p_1+q_1-p_2-q_2-\sum k_{j_1}-\sum k'_{j_2})+
D_\rQCED} 
{\tilde{\bar\beta}_{n,m}(k_1,\ldots,k_n;k'_1,\ldots,k'_m)}
,
\label{subp15b}
\end{eqnarray}
\small
where {\em new} (YFS-style) {\em non-Abelian} residuals 
{$\tilde{\bar\beta}_{n,m}(k_1,\ldots,k_n;k'_1,\ldots,k'_m)$} have {$n$} 
hard gluons and {$m$} hard photons. Definitions of the infrared functions 
${\rm SUM_{IR}(QCED)}$ and ${ D_\rQCED}$ and of the residuals are given in Ref.~\cite{mcnlo-hwri}. In the framework of shower/ME matching,  we have the replacements {$\tilde{\bar\beta}_{n,m}\rightarrow \hat{\tilde{\bar\beta}}_{n,m}$}. These replacements allow us, via the basic formula
\begin{equation}
{d\sigma} =\sum_{i,j}\int dx_1dx_2{F_i(x_1)F_j(x_2)} d\hat\sigma_{\rm res}(x_1x_2s),
\label{bscfrla}
\end{equation}
to proceed with connection to MC@NLO~\cite{mcnlo} as explained in Ref.~\cite{mcnlo-hwri}, to MG5\_aMC@NLO~\cite{mg5aMC} as explained in Refs.~\cite{bsh}, 
or to MG5\_aMC@NLO as explained in 
Ref.~\cite{liu-ward}, or to KKMChh with a Herwig6.5~\cite{hwg} or 
Herwiri1.031~\cite{hwri} 
shower as explained in Refs.~\cite{kkmchh23,kkmchh4,rdcr17}.\par
\section{New Results and New Issues in LHC/FCC Physics}
IR-improved results for precision LHC studies, such 
as those in Refs.~\cite{mcnlo-hwri,bsh} where significant improvement 
is seen the data and the IR-improved theory, show that, 
for example, in the W $+\ge$3 jets and W$ +\ge$1 jet production at the LHC 
for the third leading jet $p_T$, the improved regime extends out to 100 GeV$/c$
at 8 TeV cms energies. The robustness of this regime is new.\par
In the context of the precision studies of the single Z production at the 
LHC, {\KK}MChh with Herwig6.5 or Herwiri1.031 carry the CEEX resummed exact 
${\cal O}(\alpha^2 L)$  corrections
in which the photon transverse degrees of freedom are treated exactly. 
We see evidence of the importance of these degrees of freedom to precision
photonic ISR in Ref.~\cite{kkmchh4}.

\begin{figure}[ht]
\setlength{\unitlength}{\textwidth}
\begin{picture}(1,0.35)
\put(0,0){\includegraphics[width=0.48\textwidth]{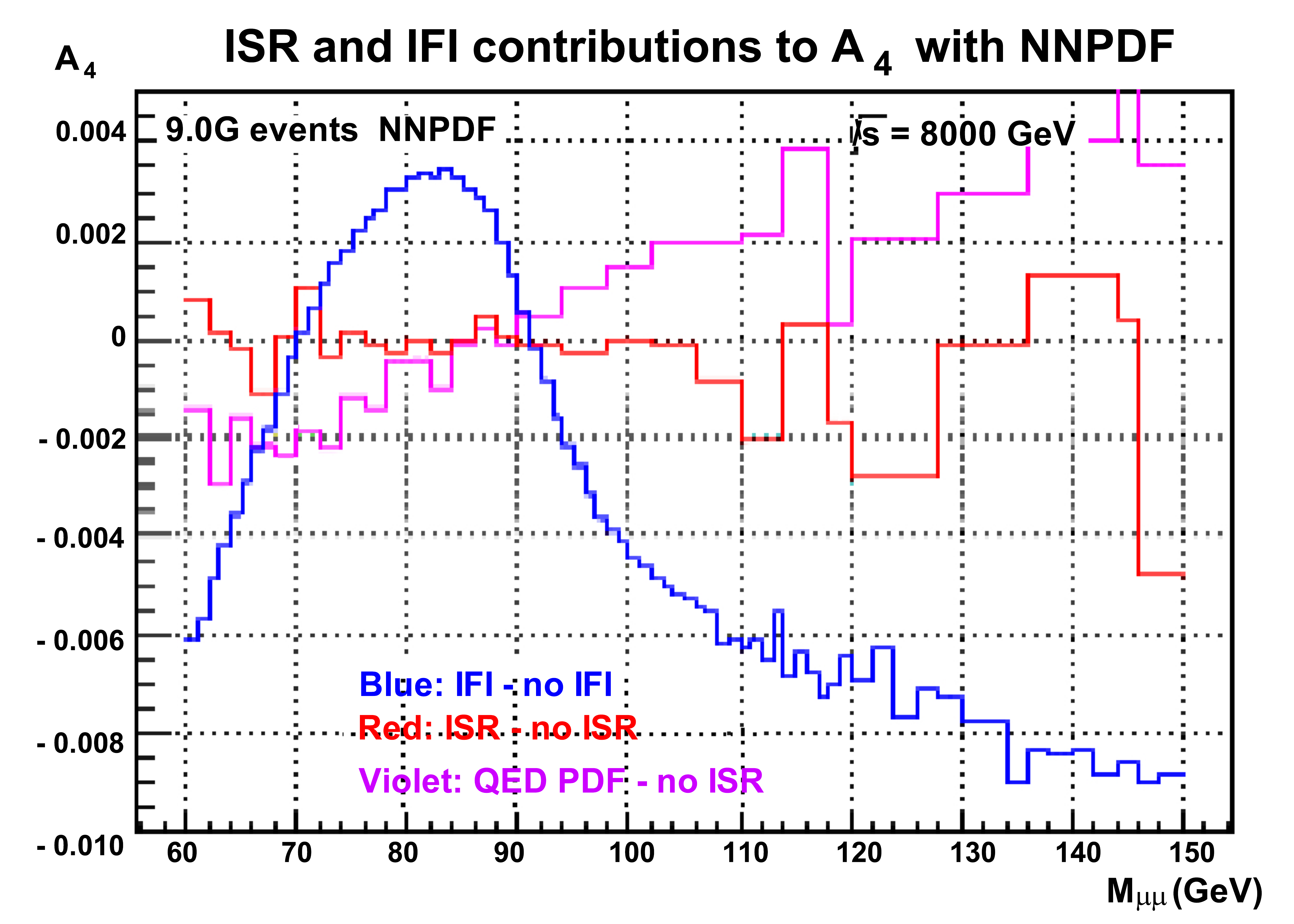}}
\put(0.48,0){\includegraphics[width=0.48\textwidth]{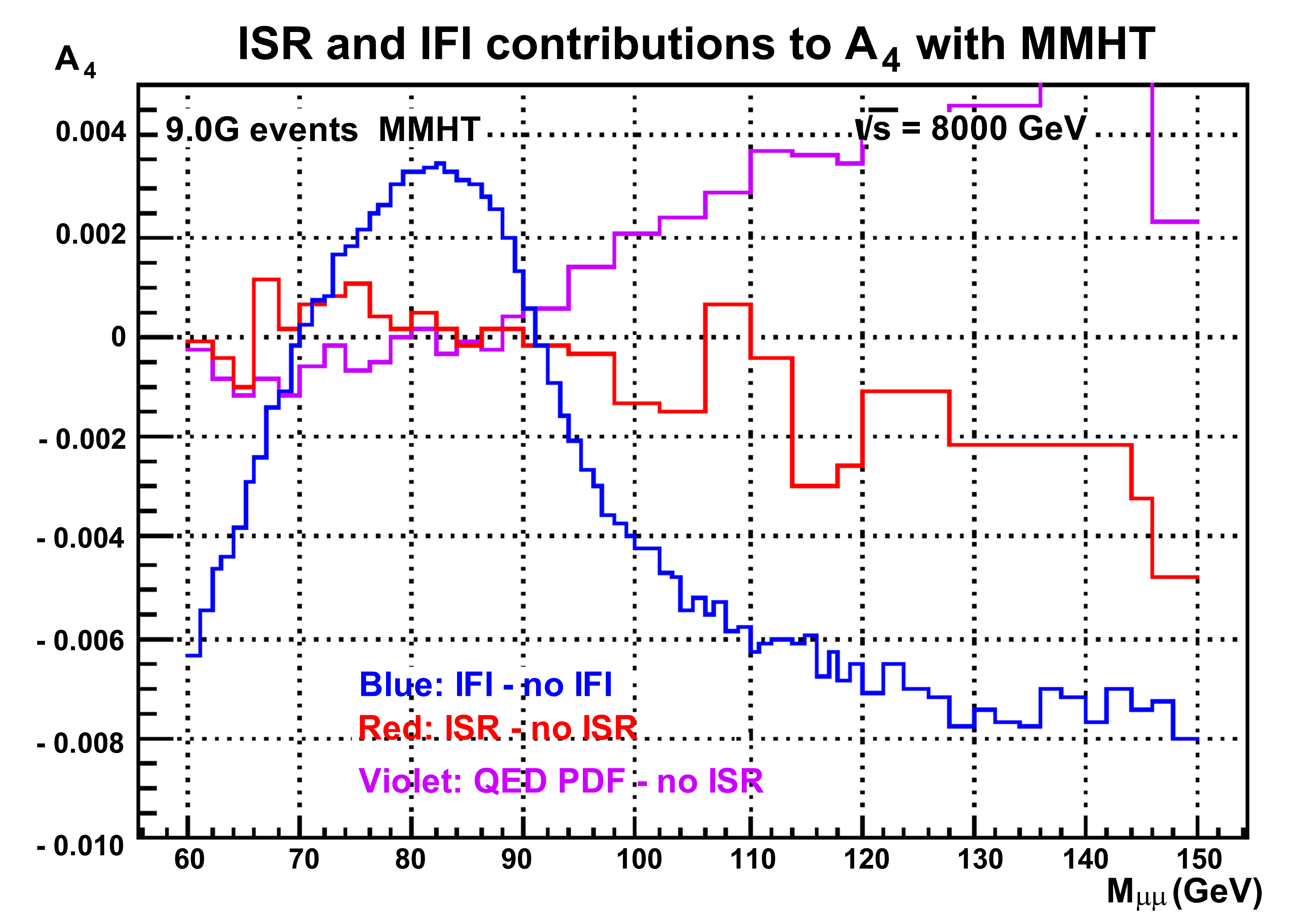}}
\end{picture}
\caption{ISR and IFI and QED vs normal PDF comparisons for 
$A_4$(calculated as $\frac{8}{3}A_\mathrm{FB}$) as a function of 
$M_{\mu\mu}$ without lepton cuts. The left figure uses 
NNPDF3.1 NLO~\cite{nnpdf3.1} and the right
figure uses MMHT2014 NLO~\cite{mmht2014}.}
\label{fig1}
\end{figure}
KKMChh's ISR corrections to forward-backward asymmetry 
shown in Fig.~1 differ from the effect of simply switching from a standard
to QED-corrected PDF set.
Issues such as the amount of QED in non-QED PDFs with starting scale 
$Q_0\cong 2$ GeV affect this comparison.  The matching of KKMChh's ISR to a 
QED-corrected PDF set is addressed by introducing ``Negative ISR''
(NISR) as explained in Ref.~\cite{say-ichep22,sj-to-appear},
where we find that adding KKMChh's ISR directly to a non-QED NNPDF3.1
PDF set, closely reproduces the effect of using a 
QED-corrected NNPDF3.1 set with NISR matching in KKMChh. 

The IR-improvement of precision phenomenology also 
extends to the FCC, to both the FCC-ee~\cite{bflw-ichep22} and the FCC-hh. 
For the latter, we note the result in Ref.~\cite{rdcr17} that, in probing the 
discovery reach for single Z production in the $p_\mathrm{T,Z} > 1$ TeV regime,
the IR-improvement does not compromise that reach. 
The outlook for the IR-improvement of precision physics at the LHC, FCC, and other planned energy frontier colliders, such as CEPC~\cite{cepc}, CPPC~\cite{cppc}, CLIC~\cite{clic-ilc}, and ILC~\cite{clic-ilc}, is therefore a rich one.
\par 
\acknowledgments
We acknowledge computing support from the Institute of Nuclear Physics, PAN, 
Krakow. S. Y. acknowledges support from The Citadel Foundation.
\par

\end{document}